\documentclass[letter,traditabstract]{aa} 
\usepackage{graphicx}
\usepackage{txfonts}
\def  \cmsq     {\ifmmode {\rm cm}^{-2} \else cm$^{-2}$\fi}
\def  \ergs     {\ifmmode {\rm erg\,s}^{-1} \else erg s$^{-1}$\fi}
\def  \ergcms   {\ifmmode {\rm erg\,cm}^{-2}\,{\rm s}^{-1}
                        \else erg\,cm$^{-2}$\,s$^{-1}$\fi}
\def \lhard  {\ifmmode {\rm L_{2-10keV}} \else ${\rm L_{2-10keV}}$\fi}
\def \nh  {\ifmmode {\rm N_{H}} \else ${\rm N_{H}}$\fi}
\def \Msun {\ifmmode M_{\odot} \else $M_{\odot}$\fi}
\def \Lsun {\ifmmode L_{\odot} \else $L_{\odot}$\fi}
\def \spitzer  {{\it Spitzer}}
\def \chandra  {{\it Chandra}}

\def \herschel {{\it Herschel}}

\begin{document}
   \title{Star formation in AGN hosts in GOODS-N\thanks{Herschel is an 
  ESA space observatory with science instruments provided by European-led 
  Principal Investigator consortia and with important participation from 
  NASA.} }

   \subtitle{}

   \authorrunning{Shao et al.}
   \titlerunning{Star formation in GOODS-N AGN}

   \author{L.~Shao\inst{1}
           \and D.~Lutz\inst{1}
           \and R.~Nordon\inst{1}
           \and R.~Maiolino\inst{2}
           \and D.M.~Alexander\inst{3}
           \and B.~Altieri\inst{4}
           \and P.~Andreani\inst{5,6}
           \and H.~Aussel\inst{7}
           \and F.E.~Bauer\inst{8}
           \and S.~Berta\inst{1}
           \and A.~Bongiovanni\inst{9,10}
           \and W.N.~Brandt\inst{11}
           \and M.~Brusa\inst{1}
           \and A.~Cava\inst{9,10}
           \and J.~Cepa\inst{9,10}
           \and A.~Cimatti\inst{12}
           \and E.~Daddi\inst{7}
           \and H.~Dominguez-Sanchez\inst{12}
           \and D.~Elbaz\inst{7}
           \and N.M.~F\"orster~Schreiber\inst{1}
           \and N.~Geis\inst{1}
           \and R.~Genzel\inst{1}
           \and A.~Grazian\inst{2}
           \and C.~Gruppioni\inst{12}
           \and G.~Magdis\inst{7}
           \and B.~Magnelli\inst{1}
           \and V.~Mainieri\inst{5}
           \and A.M.~P\'erez~Garc\'ia\inst{9,10}
           \and A.~Poglitsch\inst{1}
           \and P.~Popesso\inst{1}
           \and F.~Pozzi\inst{12}
           \and L.~Riguccini\inst{7}
           \and G.~Rodighiero\inst{13}
           \and E.~Rovilos\inst{1}
           \and A. Saintonge\inst{1}
           \and M.~Salvato\inst{14}
           \and M.~Sanchez~Portal\inst{4}
           \and P.~Santini\inst{2}
           \and E.~Sturm\inst{1}
           \and L.J.~Tacconi\inst{1}
           \and I.~Valtchanov\inst{4}
           \and M.~Wetzstein\inst{1}
           \and E.~Wieprecht\inst{1}
           }

\institute{See online appendix for author affiliations}           

   \date{received 31 March 2010  ; accepted 13 May 2010}

\abstract
{Sensitive \herschel\ far-infrared observations can break 
degeneracies that were inherent to previous studies of star formation in high-z
AGN hosts. Combining PACS 100 and 160$\mu$m observations of the GOODS-N 
field with 2Msec \chandra\ data, we detect $\sim$20\% of X-ray AGN 
individually at $>3\sigma$.
The host far-infrared luminosity of AGN with $\lhard\approx 10^{43}\ergs$ 
increases
with redshift by an order of magnitude from z=0 to z$\sim$1. In contrast, 
there is little dependence of far-infrared luminosity on AGN luminosity, for 
$\lhard\lesssim 10^{44}\ergs$ AGN at z$\gtrsim$1. We do not 
find a dependence of far-infrared luminosity on X-ray obscuring column,
for our sample which is dominated by $\lhard<10^{44}\ergs$ AGN.
In conjunction with properties
of local and luminous high-z AGN, we interpret these results as reflecting the 
interplay between two paths of AGN/host coevolution. A correlation of AGN
luminosity and host star formation is traced locally over a wide range of 
luminosities and also extends to luminous high z AGN. This correlation
reflects an evolutionary connection, likely via merging.
For lower AGN luminosities, star formation is similar to that in non-active 
massive galaxies and shows little dependence on AGN luminosity. 
The level of this secular, non-merger driven star formation increasingly 
dominates over the correlation at increasing redshift. 
}
 


   \maketitle
%

\section{Introduction}

Measuring the star formation rate of the host galaxy is important for
studying the co-evolution of active galactic nuclei (AGN) and their hosts.
It is often difficult because the AGN can outshine the host at many
wavelengths. 
However, the rest frame far-infrared/submm emission 
appears dominated by the host for AGN with 
$\rm\nu L_\nu$(60$\mu$m)$\approx$0.1\ldots 0.2 $\rm L_{Bol,AGN}$ and higher
(e.g. Netzer et al. \cite{netzer07} and introduction to Lutz et al. 
\cite{lutz10}, see also Wang et al. \cite{wang08} for luminous high-z QSOs)
and has been used as a host star formation rate diagnostic of high-z AGN
(e.g. Serjeant \& Hatziminaoglou \cite{serjeant09},
Mullaney et al. \cite{mullaney10}, and Lutz et al. \cite{lutz10}).
\herschel\ can detect much lower star formation rates than previous 
far-infrared and submm studies. It further improves such work by measuring 
the rest frame far-infrared SED peak without extrapolation from longer 
wavelengths.
We here present a first \herschel\ study of rest frame far-infrared
emission and host star formation in X-ray selected AGN in the GOODS-N field. 
Throughout the paper, we adopt an $\Omega_m =0.3$, $\Omega_\Lambda =0.7$ and
$H_0=70$ km\,s$^{-1}$\,Mpc$^{-1}$ cosmology.

\section{Data}
We use the v2.2 100 and 160\,$\mu$m maps of the GOODS-N field obtained 
with PACS (Poglitsch et al. \cite{poglitsch10}) on board \herschel\ 
(Pilbratt et al. \cite{pilbratt10}) as a 
science demonstration project for the PACS Evolutionary Probe (PEP) 
guaranteed time survey. We use the PACS catalog extracted with 
IRAC/MIPS24$\mu$m based position priors by B. Magnelli, to a 3$\sigma$ depth 
of $\sim$3.0 mJy and $\sim$5.7 mJy at  100 and 160$\mu$m, respectively. 
Since there are \chandra\ sources that are not detected at 24$\mu$m, we 
have verified by individual comparison to a blind catalog that we are not 
missing PACS 
detections on such sources when using the 24$\mu$m prior catalog.
For samples of objects undetected by PACS we stack, using the X-ray 
positions, into the residual 
maps obtained after subtracting the sources in the prior catalog from 
the original maps. See also the appendix to 
Berta et al. (\cite{berta10}) for a description of the data and catalogs.

The CDFN 2 Msec \chandra\ X-ray catalog of Alexander et al. 
(\cite{alexander03}) provides the basis for our AGN selection. 
328 of its 503 X-ray sources lie within the part of the PACS map, 
which has at least half of the coverage of the central homogeneously covered
region; we restrict ourselves to this subset. 
Reaching $\approx 2.5\times 10^{-17}\ergcms$ in the 0.5--2keV band, CDFN 
X-ray detections include a noticeable  number of star forming galaxies
without significant AGN contributions, in addition to bona-fide AGN.
Bauer et al. (\cite{bauer04}) have outlined for this very sample a 
combination of criteria
to distinguish X-ray AGN from star formation dominated objects, based on
X-ray luminosity, X-ray obscuring column or hardness, optical spectroscopic
classifications, and X-ray/optical flux ratio. We use an updated version of 
their classifications 
for the GOODS-N X-ray sources, which separates the
328 sources with good PACS coverage into 224 X-ray AGN, 67 X-ray detected 
star forming galaxies, and 37 other targets (stars or unidentified).
We use both spectroscopic (Barger et al. \cite{barger08}, 57\% of our 
X-ray AGN) and photometric redshifts 
(F. Bauer, mostly based on Barger et al. \cite{barger03}).
X-ray spectral fitting (F. Bauer et al., in prep.) provides intrinsic 
X-ray luminosities \lhard\ and obscuring columns \nh\ for the X-ray sources. 
The fits use absorbed powerlaws with a fixed Galactic column of 
1.6$\times 10^{20}$\,cm$^{-2}$ and a variable obscuring column at the redshift
of the source. The photon index is allowed to vary for sources above 
150 net counts in the 0.5-8keV band, and is fixed to 1.85 below.

\section{Results}

The far-infrared detection rate of the 328 X-ray sources above the 3$\sigma$ 
level in at least one of the two PACS bands is 28\%\ (Table~\ref{tab:detrate}).
For the three subgroups, it is 60\%\ (galaxies), 21\% (AGN) and 14\%\ (other).
The large detection rate of `galaxies' is consistent with their
X-ray emission arising from star formation related processes. A detailed 
discussion will be provided elsewhere. In the following, we focus on the
224 X-ray AGN only.

\begin{table}
\caption{PACS detections of GOODS-N X-ray sources}       
\label{tab:detrate}
\centering                       
\begin{tabular}{lcccc}       
\hline\hline            
Class   &Sources&\multicolumn{3}{c}{$>3\sigma$ detections at}\\
&&100$\mu$m&160$\mu$m&100 or 160$\mu$m\\
\hline                     
All     & 328& 76& 81& 91\\
Galaxies&  67& 37& 37& 40\\
AGN     & 224& 35& 39& 46\\
Other   &  37&  4&  5&  5\\
\hline                               
\end{tabular}
\end{table}

We use both the individual detections and stacks of nondetections derived via 
the stacking library of Bethermin et al. (\cite{bethermin10}). 
The 178 AGN that are individually undetected in both PACS bands are 
statistically well detected in the stacks at 0.79$\pm$0.08mJy (100$\mu$m) and 
1.46$\pm$0.16mJy (160$\mu$m). These detection errors are derived from multiple 
stacks at random positions over the same coverage region of the PACS residual 
maps, and using the number of targets as the sample of interest.

To derive far-infrared (FIR) luminosities with minimal assumptions about SED 
shapewe compute the rest frame 60$\mu$m luminosity $\nu$L$_\nu$(60$\mu$m) 
using 
the detection wavelength closer to rest 60$\mu$m or log-linearly interpolating 
for detections in both bands at 0.67$<$z$<$1.67. 
We treat stacked 100$\mu$m and 160$\mu$m fluxes equivalently to derive 
the mean rest frame 60$\mu$m luminosities of the individually
undetected sources. Given the strong 
K-corrections for PACS fluxes over the redshift range of X-ray AGN, this 
requires restricted redshift ranges for the stacks, we limit to bins with
$\Delta z/(1+z)<0.4$ and adopt the median z of the particular sample. 
Luminosities for the combined sample of detections and nondetections
are obtained as averages weighted by the number of sources. 

\begin{figure}
\centering
\includegraphics[width=0.96\columnwidth]{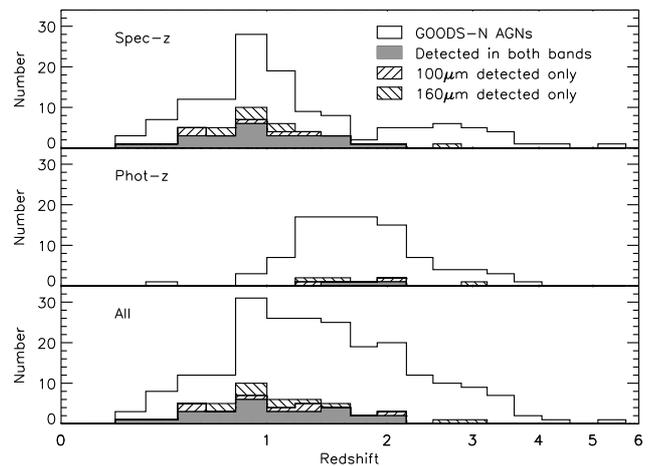}
\caption{Redshift distribution of the 224 GOODS-N X-ray AGN in the region
with good PACS coverage. 46/224 are individually detected in at least
one of the PACS bands.}
\label{fig:redshiftdist}
\end{figure}

\begin{figure}
\centering
\includegraphics[width=0.96\columnwidth]{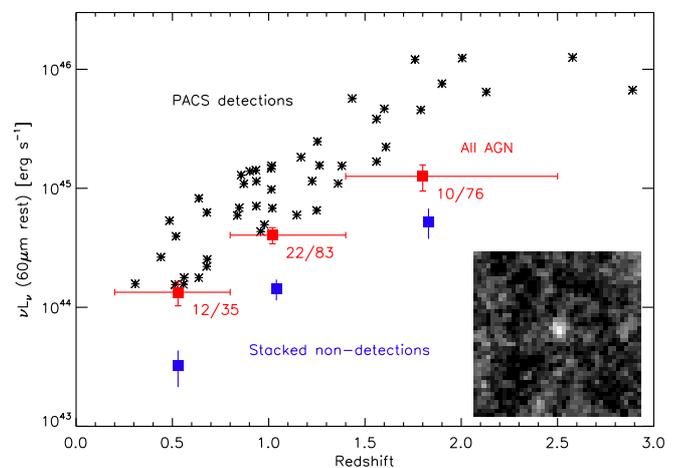}
\caption{Far-infrared luminosities of GOODS-N X-ray AGN as a function of 
redshift. Symbols labelled `All AGN', with the redshift range indicated,
average the detections and nondetections. Their uncertainty
is derived from bootstrapping into the combined sample. Number of detections 
and total number of sources in each bin are indicated. The insert shows 
the stack for the indvidually undetected 0.8$<$z$<1.4$ sources.}
\label{fig:ztrend}
\end{figure}

Figure~\ref{fig:redshiftdist} shows the redshift distribution of all X-ray 
AGN in 
our sample. As expected, PACS photometry is most efficient in detecting 
z$\lesssim$2.5 AGN hosts. We also focus on z$\leq$2.5 in order to probe 
far-infrared rest frame $>45\mu$m wavelengths, beyond the mid-infrared that 
is more easily AGN dominated. Table~\ref{tab:ztrend} and Fig.~\ref{fig:ztrend}
show 60$\mu$m luminosities
as a function of redshift, separately for the PACS detections,
stacks of nondetections and the combined sample. Host 60$\mu$m 
luminosities increase with redshift. An increase is seen in detections, 
stacked nondetections, and in the averages for the combined sample. 
This trend cannot be simply due to the increase of FIR detection 
limit with redshift, which would leave the average luminosities from  
the combination of detections and nondetections unchanged. 

\begin{table}
\caption{Mean FIR luminosities of different AGN groups} 
\label{tab:ztrend}
\centering                       
\begin{tabular}{cccrrr}       
\hline\hline            
z&\lhard&$\rm N_{Det}/N$&Detected&Stack&All\\
&\ergs &&\multicolumn{3}{c}{$\nu$L$_\nu$(60$\mu$m rest), 10$^{44}$ erg s$^{-1}$}\\
\hline                     
0.2--0.8&all          &12/35  & 3.3&0.32& 1.33$\pm$0.31\\
0.8--1.4&all          &22/83  &11.3&1.43& 4.04$\pm$0.62\\
1.4--2.5&all          &10/76  &61.1&5.27&12.61$\pm$3.12\\ \hline
0.2--0.8&$<10^{42}$   &4/16   & 1.9&0.09& 0.55$\pm$0.23\\
0.8--1.4&$<10^{42}$   &1/9    & 6.9&1.40& 2.01$\pm$0.91\\
1.4--2.5&$<10^{42}$   &0/1    &    &3.92& \\ \hline
0.2--0.8&$10^{42-43}$ &2/12   & 4.4&0.60& 1.23$\pm$0.51\\
0.8--1.4&$10^{42-43}$ &7/37   &12.3&1.56& 3.60$\pm$0.73\\
1.4--2.5&$10^{42-43}$ &6/18   &63.4&4.35&24.04$\pm$8.45\\ \hline
0.2--0.8&$10^{43-44}$ &6/7    & 3.8&1.91& 3.54$\pm$0.89\\
0.8--1.4&$10^{43-44}$ &13/32  &10.5&1.61& 5.22$\pm$1.11\\
1.4--2.5&$10^{43-44}$ &3/51   &69.3&5.30& 9.06$\pm$2.80\\ \hline
0.2--0.8&$>10^{44}$   &0/0    &    &    &\\
0.8--1.4&$>10^{44}$   &1/5    &18.2&2.32& 5.50$\pm$3.56\\
1.4--2.5&$>10^{44}$   &1/6    &22.2&8.58&10.85$\pm$4.77\\ \hline
\hline
z&$\rm N_H$&$\rm N_{Det}/N$&Detected&Stack&All\\
        &cm$^{-2}$&&\multicolumn{3}{c}{$\nu$L$_\nu$(60$\mu$m rest), 10$^{44}$ erg s$^{-1}$}\\
\hline    
0.8--1.4&$<10^{22}$  & 2/10&13.7&1.12&4.15$\pm$1.25\\
0.8--1.4&$10^{22-23}$& 6/25&11.6&1.47&3.89$\pm$0.91\\
0.8--1.4&$10^{23-24}$&10/28& 9.6&1.76&4.57$\pm$0.89\\
0.8--1.4&$>10^{24}$  &  0/5&    &1.67&1.67$\pm$0.88\\
\hline                               
\end{tabular}
Note: Average FIR luminosities are given separately for the individual
$>3\sigma$ detections, the stack of the nondetections, and the 
number-weighted mean for all sources. Errors for luminosities are standard  
deviations from bootstrapping and are dominated by the dispersion of the AGN 
population rather than measurement error. 
\end{table}

\begin{figure}
\centering
\includegraphics[width=0.98\columnwidth]{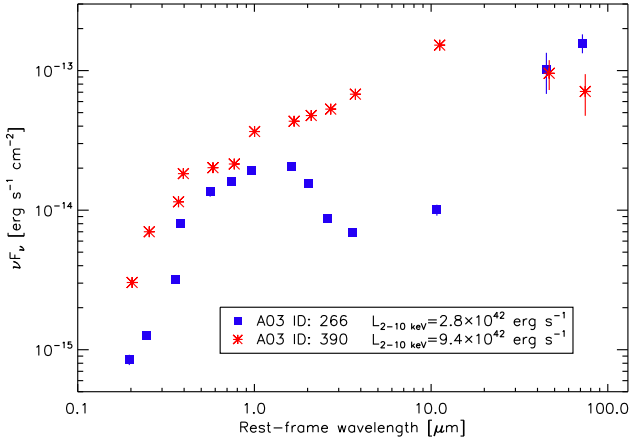}
\caption{Examples of z$\sim$1.2 AGN illustrating the range of optical to 
far-infrared SEDs.
ID 390 (Alexander et al. \cite{alexander03}) is dominated by AGN continuum 
over the optical to mid-infrared, while
ID 266 which hosts an high obscuring column AGN is dominated by an 
optical/NIR stellar bump and a pronounced far-infrared peak.}
\label{fig:exampleseds}
\end{figure}

Mean fluxes for the PACS $>3\sigma$ detections and stacked mean for the 
nondetections differ by about an order of magnitude in each of the three 
redshift bins. Such a large difference must reflect a wide intrinsic 
distribution of far-infrared luminosities. For the simplifying assumption
of a log-normal distribution of far-infrared luminosities, the ratio 
of mean detections to the stacked mean as well as the $\sim$20\%\ detection 
rates can be reproduced with an intrinsic dispersion of about 0.5 dex for the 
distribution of far-infrared luminosities in each bin. The detailed shape of 
this distribution is not constrained but must be wide. The mean 
160$\mu$m flux for all the 1.4$<$z$<$2.5 AGN is 3mJy (in agreement with 
tentative MIPS
values for ECDFS AGN by Papovich et al. \cite{papovich07}). Given the wide 
distribution of FIR luminosities, the typical (median) flux must be lower, 
by a factor $\sim$2 for the log-normal distribution. A considerable variety 
exists not only in far-infrared
luminosity but also in mid-to far-infrared SED shape 
(Fig.~\ref{fig:exampleseds}), from SEDs quite similar to those of star
forming galaxies to ones that are clearly AGN dominated over a wide 
wavelength range.  
 
\begin{figure}
\centering
\includegraphics[width=0.96\columnwidth]{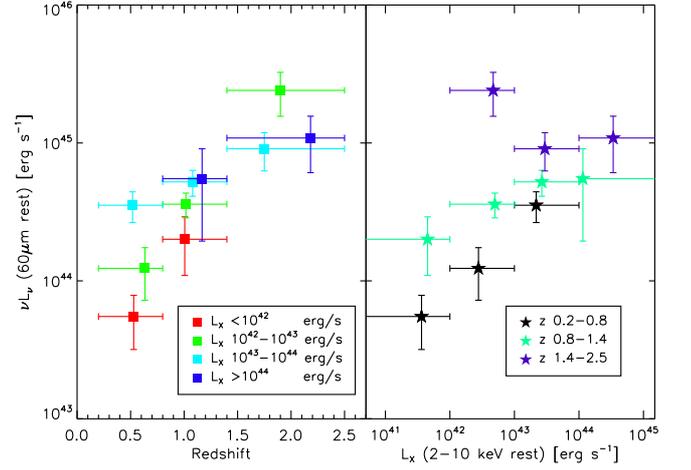}
\caption{Left: Far-infrared luminosity as a function of 
redshift, for different bins in intrinsic rest frame 2-10keV X-ray 
luminosity. Values
reflect the mean of detections and nondetections, and errors are based
on bootstrapping into the respective sample. Right: FIR luminosity as a 
function of intrinsic hard X-ray luminosity, for different redshift bins}
\label{fig:trendgrid}
\end{figure}

\begin{figure}
\centering
\includegraphics[width=0.96\columnwidth]{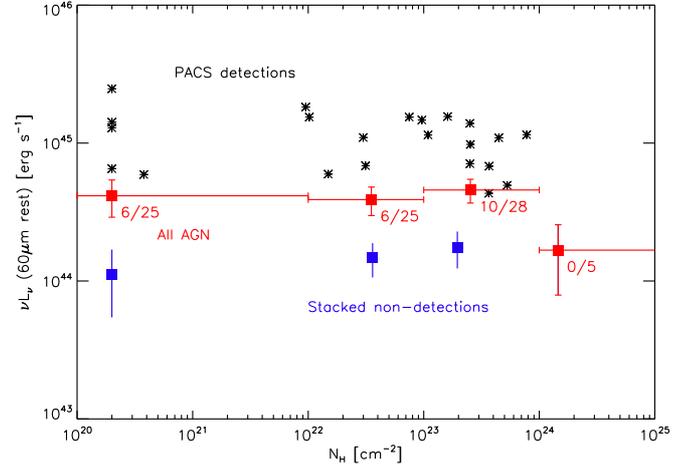}
\caption{Far-infrared luminosity as a function of X-ray obscuring column.
Sample for redshifts 0.8$<$z$<$1.4. Low obscuring column objects have been 
placed at $2\times 10^{20}$cm$^{-2}$. Symbols are as in Fig.~\ref{fig:ztrend}.}
\label{fig:nh}
\end{figure}

The increase of host far-infrared emission with redshift in 
Fig.~\ref{fig:ztrend} could still be influenced by the increase of AGN 
luminosity
with redshift that is inherent to this X-ray selected sample. To break possible
degeneracies, we have binned our sample in 3$\times$4 bins defined by redshift
(z=0.2--0.8, 0.8--1.4, 1.4--2.5) and by intrinsic hard X-ray luminosity
(\lhard$<10^{42}$\ergs, $10^{42}$--$10^{43}$, $10^{43}$--$10^{44}$, 
$>10^{44}$).  Table~\ref{tab:ztrend} lists FIR luminosities for these bins,
for the figures in the  following discussion we omit bins with less than 5
objects and correspondingly large errors. Errors on the average
FIR luminosity  are dominated by the 
variations in the underlying population rather than by photometric error.  
For that reason, errors in Table~\ref{tab:ztrend} and the figures are 
standard deviations derived from bootstrap estimates for the respective 
subsamples.

Fig~\ref{fig:trendgrid}~(left) shows that the increase of host 
FIR luminosity 
with redshift is clearly preserved when considering AGN luminosity bins 
separately. Focusing on the far-infrared luminosities of 
\lhard=$10^{42}$--$10^{43}$ AGN, for which there are 
more than 10 objects in each  redshift bin, there is no overlap in the 
99\% confidence intervals of FIR luminosity comparing the z=0.2--0.8 and the z=1.4--2.5 redshift range 
($<2.8$ vs. $>6.1\times 10^{44}\ergs$). These 99\% confidence intervals were 
directly 
derived by bootstrapping and thus include non-gaussianity of distributions
of individual source properties or of errors.
The higher the AGN luminosity bin we consider, the higher is the FIR 
luminosity at low z but then its increase with redshift less steep.
Conversely, when studying FIR luminosity as a function of AGN luminosity 
(Fig.~\ref{fig:trendgrid}~right) there is an increase
with AGN luminosity in the lowest redshift bin that flattens at higher
redshift, with no significant trend left at z$>$1.4 .

In the luminosity range covered by our sample, we do not find a significant 
trend of FIR luminosity with X-ray obscuring 
column (Fig.~\ref{fig:nh}). Such a trend would be expected in merger 
evolutionary scenarios 
that are invoking a  sequence starburst - obscured AGN - unobscured AGN
(e.g. Sanders et al. \cite{sanders88}, Hopkins et al. \cite{hopkins06}). 

\section{Evolution of the relation between AGN luminosity and host star 
formation}

\begin{figure}
\centering
\includegraphics[width=\columnwidth]{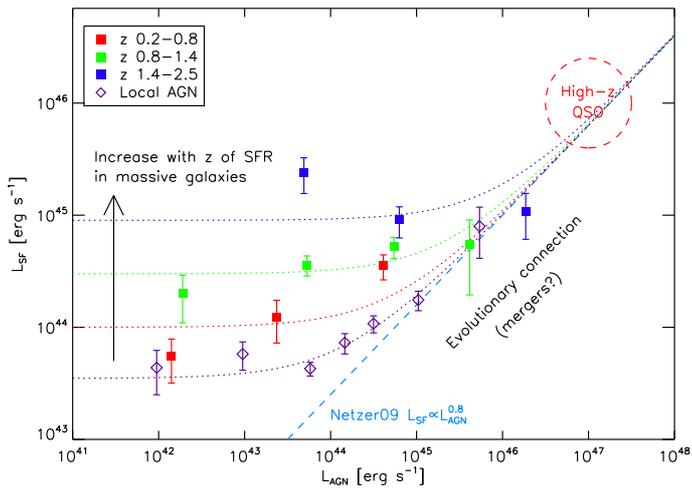}
\caption{Star forming (=far-infrared) luminosity vs. AGN luminosity for the
GOODS-N AGN and a local reference sample of extremely hard X-ray selected
BAT AGN. The dotted colored lines indicate schematically how the observations
are explained by the combination of a diagonal `evolutionary connection' 
trend with a general increase of host star formation with redshift in hosts
of moderate luminosity AGN, similar to that for the general galaxy population.
The dashed line is the relation implied by Netzer et al. (\cite{netzer09}).}
\label{fig:growth}
\end{figure}

Due to the excellent sensitivity of \herschel\/PACS to rest frame far-infrared 
emission in the hosts of z=0.2--2.5 X-ray selected AGN, we have been able to 
break the redshift/luminosity degeneracy that affected previous submm-based 
studies of AGN host star formation (e.g. Lutz et al. \cite{lutz10}). We
avoid having to make SED assumptions that were 
necessary for submm-based studies. Compared to the \spitzer\ 70$\mu$m-based 
study of Mullaney et al. (\cite{mullaney10}), we are probing the rest frame 
far-infrared out to z$\gtrsim$1 where 70$\mu$m data already enter the rest 
frame mid-IR with less favourable contrast between host emission and AGN 
heated dust.
Results of these submm as well as 70$\mu$m based studies agree well with our
findings within the mentioned constraints. To study evolution over a yet
wider redshift range, we supplement in the following  
our GOODS-N sample with local (z$<0.3$) Swift BAT 14-150keV extremely 
hard X-ray 
detected AGN (Cusumano \cite{cusumano09}). Here, far-infrared information 
is taken from the IRAS database, 
in a way consistent with the treatment of the GOODS-N AGN (see also 
Sect.~3.3 of Lutz et al. \cite{lutz10}).

A variety of studies have established a local correlation between AGN
luminosity and star forming luminosity (e.g. Rowan-Robinson \cite{roro95};
Netzer et al. \cite{netzer07}, \cite{netzer09}), at least in the 
luminosity regime of bright Seyferts or Quasars. High redshift QSOs appear to 
extend this relation to yet higher luminosities (e.g. Lutz et al. 
\cite{lutz08} and references therein). 
Figure~\ref{fig:growth} places our results in that context. Here we have 
converted from hard X-ray to AGN bolometric luminosity 
using eq. (5) of Maiolino et al. (\cite{maiolino07}) and a ratio 7 between 
bolometric and 5100\AA\ luminosity.
Local BAT AGN follow the correlation of AGN and star formation luminosity,
with the exception of a flattening at low AGN luminosities.   
Figure~\ref{fig:growth} shows that high luminosity AGN stay close to the 
correlation for all redshifts covered by our sample, but the host star
formation rates of lower luminosity AGN rise by about an order of magnitude 
from z=0 to z=1 and by about 1.5 orders of magnitude from z=0 to z=2.
Such a behavour can be explained by the combination of two paths of AGN growth.
On one path, AGN growth and host star formation are tightly coupled by an
evolutionary mechanism, likely merging. This will result in the diagonal 
correlation line in Fig.~\ref{fig:growth}. The other path reflects a secular 
evolution with no close coupling of AGN luminosity and galaxy-integrated 
host star formation rate. Connections between AGN and star formation 
phenomena on smaller spatial scales or different timescales might however 
be present. Here, star formation levels will be similar to those that are 
pervasive in massive galaxies at a given redshift. This
corresponds to the low-luminosity flattening of the relation in 
Fig.~\ref{fig:growth}
at a level increasing with redshift. The detailed slope of this flatter 
part is still poorly
constrained at current statistics. Bouch\'e et al. (\cite{bouche10})
have parametrized on the basis of a variety of observational studies star 
formation rates of star forming galaxies as a function of galaxy mass and
redshift. Their equation (1) corresponds to an increase in SFR from redshift
0 to the centers of our three redshift bins by  factors 3.0, 7.4, and 19 
respectively, 
consistent with the location of AGN on the flat `secular' path in 
Fig.~\ref{fig:growth}. As discussed in Lutz et al. 
(\cite{lutz10}), such a two-path scenario for AGN is also consistent with 
a variety of other properties of AGN hosts at high z. 

A limitation to the current study is the small 
$\sim 11^\prime\times 16^\prime$ field which restricts the number of
$\lhard >10^{44}\ergs$ AGN. Upcoming \herschel\ observations will provide 
results for larger
samples of such objects, where we will be able to investigate the relation
above these AGN luminosities, i.e. on the `merger path'. We might expect an 
upturn in host far-infrared luminosity which is suggested at z$\approx$1 for 
more luminous AGN (Lutz et \cite{lutz10}), or the possible relation of
host star formation to AGN obscuration (e.g. Page et al. \cite{page01}).


\begin{acknowledgements}
We thank the referee for helpful comments. PACS has been developed by a 
consortium of institutes led by MPE
(Germany) and including UVIE (Austria); KUL, CSL, IMEC (Belgium); CEA,
OAMP (France); MPIA (Germany); IFSI, OAP/OAT, OAA/CAISMI, LENS, SISSA
(Italy); IAC (Spain). This development has been supported by the funding
agencies BMVIT (Austria), ESA-PRODEX (Belgium), CEA/CNES (France),
DLR (Germany), ASI (Italy), and CICYT/MCYT (Spain).
\end{acknowledgements}

\Online
\appendix
\section{Author affiliations}

$^1$ MPE, Postfach 1312, 85741 Garching, Germany, \email{shao@mpe.mpg.de}
         
\noindent $^2$ INAF - Osservatorio Astronomico di Roma, via di Frascati 33, 
00040 Monte Porzio Catone, Italy        
         
\noindent $^3$ Department of Physics, Durham University, South Road,
Durham, DH1 3LE, UK
          
\noindent $^4$ European Space Astronomy Centre, Villafranca del Castillo, Spain
          
\noindent $^5$ European Southern Observatory, Karl-Schwarzschild-Stra\ss e 2,
85748 Garching, Germany

\noindent $^6$ INAF - Osservatorio Astronomico di Trieste, via Tiepolo 11, 
34143 Trieste, Italy
          
\noindent $^7$ IRFU/Service d'Astrophysique, B\^at.709, CEA-Saclay, 91191 
           Gif-sur-Yvette Cedex, France
         
\noindent $^8$ Pontificia Universidad Cat\'olica de Chile, Departamento de 
           Astronom\'ia y Astrof\'isica, Casilla 306, Santiago 22, Chile
          
\noindent $^{9}$ Instituto de Astrof\'isica de Canarias, 38205 La Laguna, 
Spain

\noindent $^{10}$ Departamento de Astrof\'isica, Universidad de La Laguna,
Spain
         
\noindent $^{11}$  Department of Astronomy and Astrophysics, 525 Davey Lab,
           Pennsylvania State University, University Park, PA 16802, USA
         
\noindent $^{12}$ Istituto Nazionale di Astronomia, Osservatorio Astronomico di 
           Bologna, Via Ranzani 1, I-40127 Bologna, Italy)
       
\noindent $^{13}$ Dipartimento di Astronomia, Universit\'a di Padova, 35122 Padova, 
            Italy
         
\noindent $^{14}$ Max-Planck-Institut f\"ur Plasmaphysik, Boltzmannstra\ss e 2, 
85748 Garching, Germany


\begin{thebibliography}{}

\bibitem[2003]{alexander03} Alexander, D.M., et al. 2003, \aj, 125, 383
\bibitem[2003]{barger03} Barger, A.J., et al. 2003, \apj, 126, 632
\bibitem[2008]{barger08} Barger, A.J., Cowie, L.L., Wang, W.-H. 2008, 
ApJ, 689, 687
\bibitem[2004]{bauer04} Bauer, F.E., et al. 2004, \aj, 128 ,2048
\bibitem[2010]{berta10} Berta, S., et al. 2010, \aap, this issue 
(arXiv 1005.1073) 
\bibitem[2010]{bethermin10} B\'ethermin,M., Dole, H., Beelen, A., Aussel, H.
2010, \aap, 512, 78
\bibitem[2010]{bouche10} Bouch\'e, N., et al. 2010, \apj, submitted 
(arXiv 0912.1858)
\bibitem[2009]{cusumano09} Cusumano, G., 2009, AIP conference proceedings 
1126, 104
\bibitem[2006]{hopkins06} Hopkins, P.F., Hernquist, L., Cox, T.J., 
DiMatteo, T., Robertson, B., Springel, V. 2006, \apjs, 163,1
\bibitem[2008]{lutz08} Lutz, D., et al. 2008, \apj, 684, 853
\bibitem[2010]{lutz10} Lutz, D., et al. 2010, \apj, 712, 1287
\bibitem[2007]{maiolino07} Maiolino, R., et al. 2007, \aap, 468, 979
\bibitem[2010]{mullaney10} Mullaney, J.R., Alexander, D.M., Huynh, M.,
  Goulding, A.D., Frayer, D. 2010, \mnras, 401, 995 
\bibitem[2007]{netzer07} Netzer, H., et al. 2007, \apj, 666, 806
\bibitem[2009]{netzer09} Netzer, H. 2009, \mnras, 399, 1907
\bibitem[2001]{page01} Page, M.J., Stevens, J.A.,
Mittaz, J.P.D., Carrera, F.J. 2001, Science 294, 2516
\bibitem[2007]{papovich07} Papovich, C., et al. 2007, \apj, 668, 45
\bibitem[2010]{pilbratt10} Pilbratt, G., et al. 2010, \aap, this issue 
\bibitem[2010]{poglitsch10} Poglitsch, A., et al. 2010, \aap, this issue
(arXiv 1005.1487)
\bibitem[1995]{roro95} Rowan-Robinson, M., 1995, \mnras, 272, 737
\bibitem[1988]{sanders88} Sanders, D.B., et al. 1988, \apj, 325, 74
\bibitem[2009]{serjeant09} Serjeant, S., Hatziminaoglou, E., 2009, \mnras,
  397, 265
\bibitem[2008]{wang08} Wang R., et al. 2008, \apj, 687, 848

\end{thebibliography}
\end{document}